\documentclass[12pt]{article}
\usepackage{amssymb,cite}
\usepackage{amsmath}
\usepackage{epsfig}
\allowdisplaybreaks
\begin{document}
\title{\bf Study of Viable Charged Wormhole Solutions in $f(R,G)$ Theory}
\author{M. Zeeshan Gul \thanks{mzeeshangul.math@gmail.com} and
M. Sharif \thanks {msharif.math@pu.edu.pk}\\
Department of Mathematics and Statistics, The University of
Lahore,\\
1-KM Defence Road Lahore-54000, Pakistan.}
\date{}
\maketitle

\begin{abstract}
In this paper, we investigate how charge and modified terms affect
the viability and stability of traversable wormhole geometry in the
framework of $f(R,G)$ theory, where $R$ is the Ricci scalar and $G$
is the Gauss-Bonnet term. For this purpose, we develop a shape
function through the Karmarkar condition to examine the wormhole
geometry. The resulting shape function satisfies all the necessary
conditions and establishes a connection between the asymptotically
flat regions of the spacetime. The behavior of energy conditions and
sound speed is checked in the presence of higher-order curvature
terms and electromagnetic field to analyze the existence of stable
traversable wormhole geometry. It is found that the traversable
wormhole solutions are viable and stable in this modified theory.
\end{abstract}
analysis.

\section{Introduction}

Einstein's general theory of relativity $(\mathcal{GR})$ is a
fundamental theory that provides a new understanding of gravity and
the nature of spacetime. In $\mathcal{GR}$, Einstein proposed that
gravity is not a force exerted by massive objects but rather a
spacetime curvature caused by mass and energy. According to
Einstein, massive objects like stars and planets curve the fabric of
spacetime around them and this curvature influences the motion of
other nearby objects. This theory revolutionized our understanding
of gravity and provided a new framework for describing the behavior
of objects in the presence of mass and energy. It has extensively
been tested/validated through various observations and experiments.
In 1917, Einstein introduced cosmological constant in his field
equations to accommodate the prevailing belief that the universe was
static and not expanding. However, in 1929, Hubble's discovery of
the expanding universe prompted Einstein to remove the cosmological
constant from his equations and revise them.

In the 1990s, different cosmic observations revealed that our
universe was in the expansion phase, which led physicists to revive
the idea of the cosmological constant \cite{1}. However, the value
of the cosmological constant required to explain the acceleration is
about one hundred twenty orders of magnitude smaller than the
predicted value by quantum field theory. This is known as the
cosmological constant problem. Although $\mathcal{GR}$ is successful
in many aspects but other unresolved problems in astrophysics and
cosmology have prompted the development of various extended
gravitational theories as
\begin{itemize}
\item Modified Newtonian Dynamics
\end{itemize}
This suggests that the laws of gravity are different on large scales
compared to the predictions of Newtonian physics and $\mathcal{GR}$.
It proposes to modify the gravitational force law, altering the
acceleration-mass relationship. This aims to explain the observed
discrepancies between the predicted and observed velocities of stars
in galaxies without invoking the existence of dark matter \cite{1a}.
\begin{itemize}
\item Scalar-Tensor Theories
\end{itemize}
These theories introduce a scalar field that couples to gravity and
affects the gravitational interactions. The scalar field can lead to
time and space-varying gravitational constants and can influence the
universe's expansion. These have been proposed as alternatives to
$\mathcal{GR}$ in attempts to explain the accelerating expansion and
the nature of dark energy \cite{1b}.
\begin{itemize}
\item $f(R)$ Gravity
\end{itemize}
In this approach, the gravitational action in $\mathcal{GR}$ is
modified by a function of the Ricci scalar. This modified theory
allows for deviations from $\mathcal{GR}$, especially in
high-curvature regimes such as during the early universe or near the
massive objects. This gravity explains the cosmic accelerated
expansion which is attributed to dark energy \cite{1c}.
\begin{itemize}
\item  The Lovelock Theory of Gravity
\end{itemize}
David Lovelock developed this theory in 1970, which is the
generalization of $\mathcal{GR}$ in higher dimensions. This theory
proposes that the gravitational field can be described by a set of
higher-order curvature tensors, which are constructed from the
Riemann tensor and its derivatives. The significant characteristic
of Lovelock gravity is that it reduces to $\mathcal{GR}$ in four
dimensions while offering a more comprehensive explanation of
gravity in higher dimensions \cite{2}. One of the main applications
of this gravity is the study of black holes in higher dimensions.
According to this theory, black holes exhibit distinct properties
compared to those predicted by $\mathcal{GR}$. Thus, Lovelock
gravity represents a significant theoretical framework for
comprehending the behavior of gravity in higher dimensions and holds
great importance for studying black holes and other astrophysical
phenomena. The first Lovelock scalar is the Ricci scalar, while the
Gauss-Bonnet invariant defined as
\begin{equation}\nonumber
G=R_{\xi\eta\lambda\gamma} R^{\xi\eta\lambda\gamma}+R^{2}
-4R_{\xi\eta}R^{\xi\eta},
\end{equation}
is the second Lovelock scalar \cite{3}. Here, the Ricci and Riemann
tensors are denoted by $R_{\xi\eta}$ and $R_{\xi\eta\lambda\gamma}$.
Nojiri and Odintsov \cite{4} established $f(G)$ gravity, which
provides fascinating insights into the expansion of the universe.
This theory is modified by introducing the curvature scalar in the
functional action named $f(R,G)$ theory \cite{5}. Moreover, this
modified theory explains the universe's accelerated expansion and
provides a unified description of the early and late-time cosmic
evolution \cite{6}-\cite{6f}.

The mysterious characteristics of our universe raise marvelous
questions for the research community. Hypothetical structures are
assumed to be the most controversial issue that yields the wormhole
($\mathcal{WH}$) structure. A $\mathcal{WH}$ is a hypothetical
concept that refers to a shortcut or a tunnel through spacetime. The
idea of a $\mathcal{WH}$ arises from the equations of
$\mathcal{GR}$, which describes the curvature of spacetime caused by
mass and energy. The two main types of $\mathcal{WH}$s are
inter-universe $\mathcal{WH}$s and intra-universe $\mathcal{WH}$s.
Inter-universe $\mathcal{WH}$s will connect different universes if
they exist, while intra-universe $\mathcal{WH}$s would connect
different regions of the same universe. The concept of
$\mathcal{WH}$s was initially proposed by the physicist Flamm
\cite{6g} in 1916, who used the Schwarzschild solution to develop
the idea of $\mathcal{WH}$ structure. Later, the concept of the
Einstein-Rosen bridge described a tunnel-like structure connecting
two separate regions of spacetime  \cite{6h}.

According to $\mathcal{GR}$, $\mathcal{WH}$s can exist if there is
enough mass and energy to warp spacetime in a specific way. They are
also believed to be highly unstable and may require exotic matter
(contradicts energy conditions) to maintain their structure. Despite
these challenges, scientists continue to explore the concept of
$\mathcal{WH}$s and their implications for our understanding of the
universe. According to Wheeler \cite{7}, Schwarzschild
$\mathcal{WH}$ solutions are not traversable due to the presence of
strong tidal forces at $\mathcal{WH}$ throat and the inability to
travel in two directions. Furthermore, the throat of $\mathcal{WH}$
rapidly expands and contracts, preventing access to anything.
However, it is analyzed that $\mathcal{WH}$s would collapse
immediately after the formation \cite{8}. The possibility of a
feasible $\mathcal{WH}$ is being challenged due to the enormous
amount of exotic matter. Thus, a viable $\mathcal{WH}$ structure
must have a minimum amount of exotic matter. Morris and Thorne
\cite{9} proposed the first traversable $\mathcal{WH}$ solution.

The most interesting subjects in gravitational physics are the study
of $\mathcal{WH}$ shape functions. Shape functions play a crucial
role in determining the properties and behavior of traversable
$\mathcal{WH}$s. They are mathematical functions that describe the
spatial geometry of a $\mathcal{WH}$, specifically the throat's
radius as a function of the radial coordinate. Different shape
functions allow us to model various types of $\mathcal{WH}$s with
distinct characteristics. One commonly used shape function is the
Morris-Thorne shape function which is employed to model spherically
symmetric $\mathcal{WH}$s \cite{10}. The choice of shape function
significantly impacts the properties of the $\mathcal{WH}$. The
non-static conformal $\mathcal{WH}$s using two different shape
functions has been investigated in \cite{11}. Recently, many
researchers \cite{13} proposed various shape functions to describe
the structure of $\mathcal{WH}$s, aiming to explore new
possibilities, characteristics and behavior of these fascinating
theoretical constructs. The development of new shape functions
enhance our understanding of $\mathcal{WH}$s and may lead to
insights into their implications for future advancements in
theoretical physics.

Various techniques have been considered to examine the
$\mathcal{WH}$ structures including the solution of metric elements,
imposing constraints on fluid parameters and employing specific
equations of state. In this context, the embedding class-I method
has been proposed to investigate cosmic objects. The static
spherically symmetric solutions are examined using the embedding
class-I condition in \cite{14}. Karmarkar \cite{15} established a
necessary constraint for static spherical spacetime that belongs to
embedding class-I. Recently, spherical objects with various matter
distributions through the embedding class-I method have been studied
in \cite{16}-\cite{20a}. The viability of traversable $\mathcal{WH}$
geometry through the Karmarkar constraint has been investigated in
\cite{21}. Numerous studies have explored the effects of charge on
cosmic structures \cite{22}. Sharif and Javed \cite{23} examined the
impact of the charge on thin-shell $\mathcal{WH}$s using the
cut-and-paste method.

Visser \cite{23a} formulated a new class of traversable
$\mathcal{WH}$s by surgically grafting two Schwarzschild spacetimes
together and found that the constructed $\mathcal{WH}$s prevent the
formation of event horizons. Ovgun \cite{23b} studied how a dark
matter medium affects the weak deflection angle of traversable
$\mathcal{WH}$s. For this purpose, he used the Gauss-Bonnet theorem
in his analysis and found that the bounce parameter influences the
weak deflection angle. Kumaran and Ovgun \cite{23c} used
Gauss-Bonnet theorem to derive the weak deflection angle for
traversable $\mathcal{WH}$s. Javed et al \cite{23d} studied the
$\mathcal{WH}$-like static aether solution and computed deflection
angle in the various mediums such as non-plasma, plasma and dark
matter. Javed et al \cite{23e} used Gibbons and Werner methods to
derive the weak deflection angle for Kalb-Ramond traversable
$\mathcal{WH}$ solutions in plasma and dark matter mediums. They
found that if one removes the effects of plasma and dark matter, the
results become identical to that of non-plasma case.

Halilsoy et al \cite{24a} used linear perturbations to analyze the
stability of the thin-shell $\mathcal{WH}$s and found that the
Hayward parameter enhances the stability of thin-shell
$\mathcal{WH}$s. Ovgun \cite{24b} developed a rotating thin-shell
$\mathcal{WH}$ through Darmois-Israel junction conditions. Richarte
et al \cite{24c} employed cut and paste technique to formulate the
traversable thin-shell $\mathcal{WH}$s. Ovgun \cite{24d} extended
the idea of Gibbons-Werner and applied the Gauss-Bonnet theorem to
the rotating/non-rotating Damour-Solodukhin $\mathcal{WH}$s to study
the weak gravitational lensing by these objects. Jusufi and Ovgun
\cite{24e} calculated the deflection angle of a rotating
$\mathcal{WH}$s by using the Gauss-Bonnet theorem. Javed et al
\cite{24f} analyzed the deflection angle of light by Brane-Dicke
$\mathcal{WH}$ in the weak field limit approximation. Ovgun
\cite{24g} obtained the evolving topologically deformed
$\mathcal{WH}$ supported in the dark matter halo and checked the
behavior of the $\mathcal{WH}$ in the inflation era.

Researchers have been exploring alternative ideas and theoretical
frameworks to understand the nature of spacetime and the possibility
of exotic structures like $\mathcal{WH}$s. The viable attributes of
$\mathcal{WH}$s provide fascinating outcomes in the framework of
modified gravitational theories. Lobo and Oliveira \cite{24}
employed equations of state and various shape functions to analyze
the structures of $\mathcal{WH}$ in $f(R)$ theory. Azizi \cite{25}
examined static spherically symmetric $\mathcal{WH}$ solutions by
considering a specific equation of state in $f(R,T)$ theory. Sharif
and Fatima \cite{26} studied traversable $\mathcal{WH}$
configuration in the framework of $f(G)$ theory. Elizalde and
Khurshudyan \cite{27} explored the viable $\mathcal{WH}$ solutions
through barotropic equation of state in $f(R,T)$ theory. Sharif and
Hussain \cite{28} investigated the viability and stability of static
spherical $\mathcal{WH}$ geometry in the realm of $f(G,T)$ gravity.
Shamir and Fayyaz \cite{29} used the same technique to formulate a
shape function in $f(R)$ theory and discovered that a small amount
of exotic matter can produce a $\mathcal{WH}$ geometry. We have
considered static spherically symmetric spacetime with a Noether
symmetry approach to examine the $\mathcal{WH}$ solutions in
$f(R,T^2)$ theory \cite{30}. Godani \cite{31} studied the viable as
well as stable $\mathcal{WH}$ solutions in $f(R,T)$ gravity. Malik
et al \cite{32} used the embedding class-I technique to study the
static spherical solutions in $f(R)$ theory. We have studied the
dynamics of gravitational collapse \cite{32a}, Noether symmetry
approach \cite{32b} and stability of the Einstein universe
\cite{32c} in $f(R,T^{2})$ gravity. Recently, Sharif and Fatima
\cite{33} have employed the Karmarkar condition to investigate the
viable $\mathcal{WH}$ structures in $f(R,T)$ theory.

This paper investigates viable traversable $\mathcal{WH}$ solutions
using the embedding class-I technique in $f(R,G)$ theory. The
analysis focuses on studying the behavior of shape function and
energy conditions in this context. Wormholes are intriguing
solutions to the Einstein field equations that have captured
significant attention due to their implications in cosmology and
interstellar travel. However, their viability and stability in the
framework of alternative gravitational theories remain an open
question. The motivation for exploring $f(R,G)$ gravity is twofold.
First, this gravitational theory is an extension of $\mathcal{GR}$
that allows for a more comprehensive description of gravitational
phenomena. Second, $\mathcal{WH}$ solutions in $f(R,G)$ gravity
offer new insights between gravity modifications and exotic
structures like $\mathcal{WH}$s. By investigating $\mathcal{WH}$s in
this modified gravity theory, we aim to contribute our understanding
on the existence and stability of $\mathcal{WH}$ solutions.
Furthermore, investigating $\mathcal{WH}$s in this framework may
shed light on the compatibility of $\mathcal{WH}$s with modified
gravity theories, which has implications for theoretical physics and
observational cosmology.

To our knowledge, there has been limited exploration of
$\mathcal{WH}$s in the context of $f(R,G)$ gravity. Our study takes
a pioneering step in examining the existence and stability of
$\mathcal{WH}$ solutions in this specific modified theory. Most
previous research on $\mathcal{WH}$s in modified theories focused on
$f(R)$ gravity or $f(G)$ gravity. In contrast, our approach
considers the joint effects of the Ricci scalar and the Gauss-Bonnet
invariant in $f(R,G)$ gravity, providing a more comprehensive
analysis. We intend to perform a detailed viability and stability
analysis of the $\mathcal{WH}$ solutions in $f(R,G)$ gravity, which
is a novel aspect of our study. By conducting a comprehensive
analysis of $\mathcal{WH}$ solutions in $f(R,G)$ gravity, our work
contributes to the broader understanding of gravitational theories
and their astrophysical implications.

The paper is structured as follows. Section \textbf{2} establishes
the field equations of charged spherical spacetime in $f(R,G)$
theory and discuss the Karmarkar condition. In section \textbf{3},
we develop the $\mathcal{WH}$ shape function using the Karmarkar
condition. We also investigate the viable traversable $\mathcal{WH}$
geometry through energy conditions. Further, we check the stability
through causality condition and Herrera cracking approach in section
\textbf{4}. The final section provides a summary of our findings.

\section{$f(R,G)$ Gravity and Karmarkar Condition}

The modified Einstein-Hilbert action in the presence of
electromagnetic field is defined as \cite{5}
\begin{equation}\label{1}
\mathcal{I}= \frac{1}{2\kappa}\int f
(R,G){\sqrt{-\mathrm{g}}}d^4x+\int
(L_{m}+L_{e})\sqrt{-\mathrm{g}}d^4x,
\end{equation}
where the matter-lagrangian and determinant of the line element are
represented by $L_{m}$ and $\mathrm{g}$, respectively. The
electromagnetic-lagrangian is expressed as
\begin{eqnarray}\label{2}
L_{e}=\varpi F_{\xi\eta}F^{\xi\eta}, \quad F_{\xi
\eta}=\varphi_{\eta,\xi}-\varphi_{\xi,\eta},
\end{eqnarray}
where $\varpi$ is an arbitrary constant and
$\varphi^{\xi}(r)=\delta^{\xi}_{0}\varphi(r)$ is the four-potential.
By varying Eq.(\ref{1}) corresponding to the metric tensor, we
obtain
\begin{eqnarray}\nonumber
R_{\xi\eta}-\frac{1}{2}\mathrm{g}_{\xi\eta}
R&=&T_{\xi\eta}+E_{\xi\eta}+\nabla_{\xi}\nabla_{\eta}f_{R}
-\mathrm{g}_{\xi\eta}\nabla_{\xi}\nabla^{\xi}f_{R}
-\frac{1}{2}\mathrm{g}_{\xi\eta}(Rf_{R}+Gf_{G}-f)
\\\nonumber
&+&2R\nabla_{\xi}\nabla_{\eta}f_{G}-2
\mathrm{g}_{\xi\eta}R\nabla_{\xi}\nabla^{\xi}f_{G}-4R^{\lambda}
_{\xi}\nabla_{\lambda}\nabla_{\eta}f_{G}
-4R^{\lambda}_{\eta}\nabla_{\lambda}\nabla_{\xi}f_{G}
\\\nonumber
&+&R_{\xi\eta}\nabla_{\xi}\nabla^{\xi}f_{G}+4\mathrm{g}
_{\xi\eta}R^{\lambda\gamma}
\nabla_{\lambda} \nabla_{\gamma}f_{G}+4R_{\xi\lambda
\eta\gamma}
\nabla^{\lambda}\nabla^{\gamma}f_{G}
\\\label{3}
&+&(1-f_{R})R_{\xi\eta}-\frac{1}{2}R\mathrm{g}_{\xi\eta}.
\end{eqnarray}
Here, $f\equiv f(G)$, $f_{G}=\frac{\partial f} {\partial G}$ and
$f_{R}=\frac{\partial f} {\partial R}$. The stress-energy tensor of
electromagnetic field is given by
\begin{equation}\label{4}
E_{\xi\eta}=\frac{1}{16\pi}\mathrm{g}_{\xi\eta}F^{\lambda
\gamma}F_{\lambda\gamma} -\frac{1}{4\pi}F^{\lambda}_{\xi}
F_{\eta\lambda}.
\end{equation}
We assume anisotropic matter distribution as
\begin{equation}\label{5}
T_{\xi\eta}=
\mathcal{U}_{\xi}\mathcal{U}_{\eta}(\varrho+P_{\bot})-P_{\bot}\mathrm{g}_{\xi\eta}
+\mathcal{V}_{\xi}\mathcal{V}_{\eta}(P_{r}-P_{\bot}).
\end{equation}
We consider static spherical metric to analyze the geometry of
$\mathcal{WH}$ as
\begin{equation}\label{6}
ds^{2}=dt^{2}e^{\mu(r)}-dr^{2}e^{\nu(r)}-d\theta^{2}r^{2}
-d\phi^{2}r^{2}\sin^{2}\theta.
\end{equation}
The Maxwell field equations are defined as
\begin{eqnarray}\label{7}
F_{\xi\eta;\lambda}=0, \quad F^{\xi\eta} _{;\eta}=4\pi J^{\xi},
\end{eqnarray}
where four-current is defined by
$J^{\xi}=\sigma(r)\mathcal{U}^{\xi}$ with $\sigma$ is the charge
density. The corresponding Maxwell field equation is
\begin{equation}\label{9}
\varphi''-\bigg(\frac{\mu'}{2}+\frac{\nu'}{2}-\frac{2}{r}\bigg)\varphi'=4\pi
\sigma e^{\frac{\mu}{2}+\nu}.
\end{equation}
Here prime=$\frac{d}{dr}$. Integrating this equation, we get
\begin{eqnarray}\label{10}
\varphi'=\frac{e^{\frac{\mu+\nu}{2}}}{r^{2}}q(r), \quad
q(r)=4\pi\int_{0}^{r}\sigma r^{2}e^{\frac{\nu}{2}}dr, \quad
E=\frac{q}{4\pi r^{2}}.
\end{eqnarray}
Using Eqs.(\ref{3}) and (\ref{6}), we obtain
\begin{eqnarray}\nonumber
\varrho&=&\frac{1}{r^{2}}(e^{\nu}+r\nu'-1)e^{-\nu}f_{R}+\frac{1}{2}
(Gf_{G}+Rf_{R}-f)-\frac{2\nu'}{r^{2}}(e^{\nu}-3)e^{-2\nu}f_{G}'
\\\label{11}
&+&\frac{1}{2r}(r\nu'-4)e^{-\nu}f_{R}'-f_{R}'' -\frac{4}{r^{2}}
(1-e^{\nu})e^{-2\nu}f_{G}''-\frac{q^{2}}{8\pi r^{4}},
\\\nonumber
P_{r}&=&\frac{1}{r^{2}}(1-e^{\nu}+r\mu')e^{-\nu}f_{R}-\frac{1}{2}
(Gf_{G}+Rf_{R}-f)-\frac{2\mu'}{r^{2}}e^{-2\nu}(e^{\nu}-3)f_{G}'
\\\label{12}
&+&\frac{1}{2r^{2}}(4r+r^{2}\mu')e^{-\nu}f_{R}'+\frac{q^{2}}{8\pi
r^{4}},
\\\nonumber
P_{\bot}&=&\frac{1}{2}
(f-Gf_{G}-Rf_{R})+\frac{1}{4r}(\mu'^{2}r-2\nu'+2r\mu''-r\mu'\nu'
+2\mu')e^{-\nu}f_{R}
\\\nonumber
&+&e^{-\nu}f_{R}''+\frac{1}{r}(\mu'^{2}+2\mu''-3\mu'\nu')e^{-2\nu}f_{G}'
+\frac{2\mu'}{r}e^{-2\nu}f_{G}''-\frac{q^{2}}{8\pi r^{4}}
\\\label{13}
&+&\frac{1}{2}(\mu'-\nu'+2r^{-1})e^{-\nu}f_{R}',
\end{eqnarray}
where
\begin{eqnarray}\nonumber
R&=&\frac{e^{-\nu}}{2r^{2}}\big[4-4e^{\nu}+r^{2}\mu'^{2}
-4r\nu'+4r\mu'-r^{2}\mu'\nu'+2r^{2}\mu''\big],
\\\nonumber
G&=&-\frac{2e^{-2\nu}}{r^{2}}\big[e^{\nu}\mu'\nu'-3\mu'\nu'
-2\mu''e^{\nu}-\mu'^{2}e^{\nu}+2\mu''+
\mu'^{2}\big].
\end{eqnarray}

Now, we use embedding class-I technique to formulate the
$\mathcal{WSF}$ that determines the $\mathcal{WH}$ geometry. The
non-vanishing components of the Riemann curvature tensor with
respect to static spherical spacetime (\ref{6}) are
\begin{eqnarray}\nonumber
R_{1212}&=&\frac{e^{\mu}(2\mu''+\mu'^{2}-\mu'\nu')}{4}, \quad
R_{3434}=\frac{r^{2}\sin^{2}\theta(e^{\nu}-1)}{e^{\nu}},
\\\nonumber
R_{1414}&=&\frac{r\sin^{2}\theta \mu'e^{\mu-\nu}}{2}, \quad
R_{2323}=\frac{r\nu'}{2}, \quad R_{1334}=R_{1224}\sin^{2}\theta.
\end{eqnarray}
The non-zero Riemann curvature tensor satisfy the Karmarkar
condition as \cite{15}
\begin{eqnarray}\label{14}
R_{1414}&=&\frac{R_{1212}R_{3434} +R_{1224}R_{1334}}{R_{2323}},\quad
R_{2323}\neq0.
\end{eqnarray}
Solving this constraint, we get
\begin{eqnarray}\label{15}
\mu'\nu'-2\mu''-\mu'^{2}=\frac{\mu'\nu'}{1-e^{\nu}}, \quad
e^{\nu}\neq1
\end{eqnarray}
Solution of this equation becomes
\begin{equation}\label{16}
e^{\nu}=ae^{\mu}\mu'^{2}+1,
\end{equation}
where integration constant is denoted by $a$.

\section{Traversable Wormhole Geometry}

We consider the Morris-Thorne spacetime to develop the
$\mathcal{WSF}$ as \cite{9}
\begin{equation}\label{17}
ds^{2}=dt^{2}e^{\mu(r)}+dr^{2}(1-hr^{-1})^{-1}
+d\theta^{2}r^{2}+d\phi^{2}r^{2}\sin\theta.
\end{equation}
Here, the shape function is denoted by $h=h(r)$ and the metric
coefficient $\mu(r)$ is defined as \cite{34}
\begin{equation}\label{18}
\mu(r)=-\frac{2b}{r},
\end{equation}
where the arbitrary constant is represented by $b$ and $\mu(r)$ is
called redshift function such that when $r\rightarrow\infty$,
$\mu(r)\rightarrow0$. Comparison of Eqs.(\ref{6}) and (\ref{17})
gives
\begin{equation}\label{19}
\nu(r)=\ln(r)-\ln(r-h(r)).
\end{equation}
Using Eqs.(\ref{16}) and (\ref{19}), we have
\begin{equation}\label{20}
h(r)=r-\frac{r^{5}}{r^{4}+4b^{2}a/e^{\frac{2b}{r}}}.
\end{equation}
For a traversable $\mathcal{WH}$ solution, the following conditions
must be satisfied \cite{9}
\begin{enumerate}
\item
$h(r)<r$,
\item
$h(r)-r=0$ at $r=r_{0}$,
\item
$\frac{h(r)-rh'(r)}{h^{2}(r)}>0$ at $r=r_{0}$,
\item
$h'(r)<1$,
\item
$\frac{h(r)}{r}\rightarrow0$ when $r\rightarrow\infty$,
\end{enumerate}
where radius of $\mathcal{WH}$ throat is defined by $r_{0}$. At
$r=r_{0}$, Eq.(\ref{20}) gives trivial solution, i.e.,
$h(r_{0})-r_{0}=0$. Therefore, we redefine Eq.(\ref{20}) for
non-trivial solution as
\begin{eqnarray}\label{21}
h(r)=r-\frac{r^{5}}{r^{4}+4b^{2}a/e^{\frac{2b}{r}}}+c.
\end{eqnarray}
Using the condition (2), we have
\begin{equation}\label{22}
h(r)=\frac{r_{0}^{4}(r_{0}-b)}{4b^{2}/e^{\frac{2b^{2}}{r_{0}}}}.
\end{equation}
Inserting this value in Eq.(\ref{21}), it follows that
\begin{eqnarray}\label{23}
h(r)=r-\frac{r^{5}}{r^{4}+r_{0}^{4}(r_{0}-c)}+c.
\end{eqnarray}
Conditions (3) and (4) are also satisfied for the specified values
of $c$. Using condition (5) in Eq.(\ref{23}), we have
\begin{equation}\label{24}
\lim _{r\rightarrow\infty}\frac{h(r)}{r}=0.
\end{equation}
Thus, the formulated shape function gives asymptotically flat
$\mathcal{WH}$ geometry. We assume $r_{0}=2$, $b=-1$ and
$c=1.6$(magenta), 1.7(orange), 1.8(red), 1.9(cyan) to analyze the
graphical behavior of the $\mathcal{WSF}$. Figure \textbf{1}
manifests that our developed shape function through Karmarkar
condition is physically viable as it satisfies all the required
conditions.
\begin{figure}
\epsfig{file=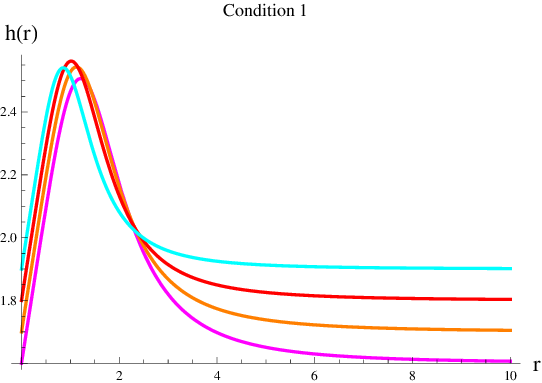,width=.5\linewidth}
\epsfig{file=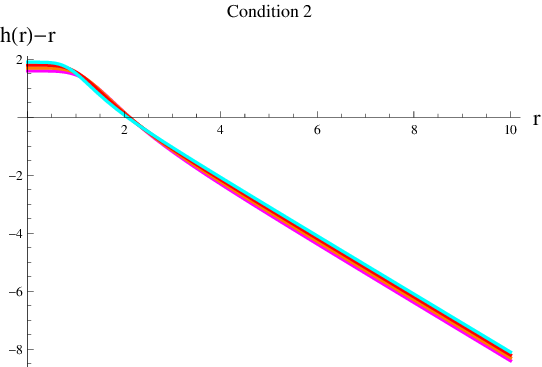,width=.5\linewidth}
\epsfig{file=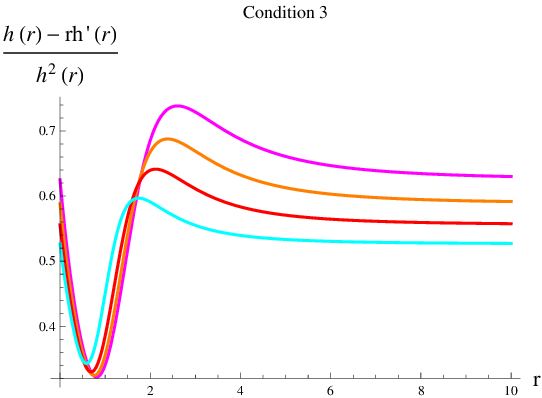,width=.5\linewidth}
\epsfig{file=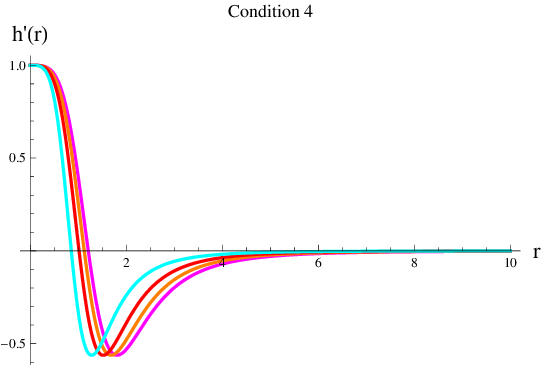,width=.5\linewidth}\center
\epsfig{file=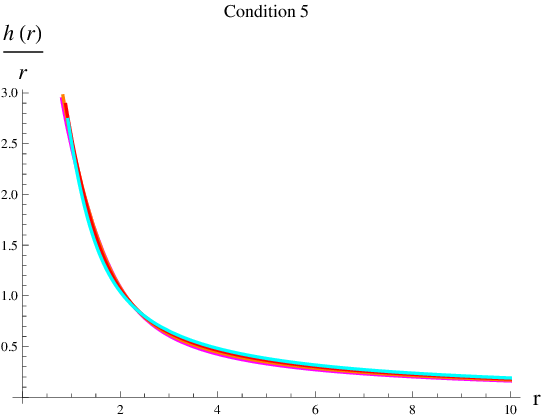,width=.5\linewidth}\caption{Behavior of
Morris-Thorne conditions corresponding to developed
$\mathcal{WSF}$.}
\end{figure}

Now, we examine how $f(R,G)$ affects the geometry of $\mathcal{WH}$.
The outcomes of our research may uncover concealed cosmological
findings on both theoretical and astrophysical levels. It is
valuable to explore alternative theories such as $f(R,G)$ to
determine the presence of hypothetical objects. This could serve as
a mathematical tool for examining various obscure features of
gravitational dynamics on a large scale. The existence of
multivariate functions and their derivatives yield field equations
(\ref{11})-(\ref{13}) in a more complex form. Hence, we cannot
deduce any result. Thus, we choose a specific model of this theory
as\cite{34a}-\cite{34b}
\begin{equation}\label{25}
f(R,G)=R+\lambda R^{2}+\alpha G^{n}+\beta G\ln(G),
\end{equation}
where $\lambda$, $\alpha$ $\beta$ and $n$ are arbitrary constants.
Since this model allows extra degrees of freedom in the field
equations, it could provide observationally well-consistent cosmic
results. The resulting equations of motion are
\begin{eqnarray}\nonumber
\varrho&=&\frac{e^{-\nu}}{r^{2}}(e^{\nu}+r\nu'-1)(1+2\lambda
R)+\frac{1}{2}G(n\alpha G^{n-1}+\beta+\beta\ln(G))
+\frac{1}{2}(\lambda R^{2}
\\\nonumber
&-&\alpha G^{n}-\beta
G\ln(G))-\frac{2\nu'}{r^{2}}(e^{\nu}-3)e^{-2\nu}G'\big\{\alpha
n(n-1)G^{n-2}+\beta G^{-1}\big\}
\\\nonumber
&+&\frac{e^{-\nu}}{r}(r\nu'-4)\lambda R'-2\lambda R''
-\frac{4}{r^{2}}e^{-2\nu}(1-e^{\nu})\big\{\alpha n(n-1)G^{n-2}G''
\\\label{27}
&+&\beta G''G^{-1}+\alpha(n-1)(n-2)nG^{n-3}G'^{2}-\beta G'^{2}G^{-2}
\big\}-\frac{q^{2}}{8\pi r^{4}},
\\\nonumber
P_{r}&=&\frac{e^{-\nu}}{r^{2}}(1-e^{\nu}+r\mu')(1+2\lambda
R)-\frac{1}{2}G(n\alpha G^{n-1}+\beta+\beta\ln(G))-\lambda R^{2}
\\\nonumber
&-&\frac{1}{2}(\alpha G^{n}+\beta
G\ln(G))-\frac{2e^{-2\nu}}{r^{2}}\mu'(e^{\nu}-3)G'\big\{\alpha
n(n-1)G^{n-2}+\beta G^{-1}\big\}
\\\label{28}
&+&\frac{e^{-\nu}}{r^{2}}(4r+r^{2}\mu')\lambda R'+\frac{q^{2}}{8\pi
r^{4}},
\\\nonumber
P_{\bot}&=&\frac{1}{2}(\alpha G^{n}+\beta
G\ln(G))-\frac{1}{2}G(n\alpha
G^{n-1}+\beta+\beta\ln(G))-\frac{1}{2}\lambda R^{2}+\frac{2\lambda
R''}{e^{\nu}}
\\\nonumber
&+&\frac{e^{-\nu}}{4r}(1+2\lambda
R)(\mu'^{2}r-2\nu'+2r\mu''-r\mu'\nu' +2\mu')+\frac{\lambda
R'}{e^{\nu}}(\mu'-\nu'+2r^{-1})
\\\nonumber
&+&\frac{e^{-2\nu}}{r}(\mu'^{2}+2\mu''-3\mu'\nu')G'\big\{\alpha
n(n-1)G^{n-2}+\beta G^{-1}\big\}
+\frac{2e^{-2\nu}}{r}\mu'\big\{\alpha n
\\\nonumber
&\times&(n-1)G^{n-2}G''+\beta
G''G^{-1}+\alpha(n-1)(n-2)nG^{n-3}G'^{2}-\beta G'^{2}G^{-2}
\big\}
\\\label{29}
&-&\frac{q^{2}}{8\pi r^{4}},
\end{eqnarray}
where
\begin{eqnarray}\nonumber
R'&=&-\frac{e^{-\nu}\nu'}{2r^2}(2r^2\mu''-4r^2\mu'\nu'+r^2\mu'^2+4r\mu'-4r
\nu'-4e^{\nu}+4)-\frac{e^{-\nu}}{r^3}
\\\nonumber
&\times&(2r^2\mu''-4r^2\mu' \nu'+r^2
\mu'^2+4r\mu'-4r\nu'-4e^{\nu}+4)+\frac{e^{-\nu}}{2r^2}(2r^2\mu'''
\\\nonumber
&-&4r^2\mu''\nu'+8r\mu''-4r^2\mu'\nu''-8r\mu'\nu'+2r \mu'^2+4
\mu'+2r^2\mu'\mu''-4\nu'
\\\nonumber
&-&4e^{\nu}\nu'-4r\nu''),
\\\nonumber
R''&=&\frac{e^{-\nu}\nu'^2}{2r^2}(2r^2\mu''-4r^2\mu'\nu'+r^2
\mu'^2+4r\mu'-4r\nu'-4e^{\nu}+4)+\frac{3e^{-\nu}}{r^4}
\\\nonumber
&\times&(2r^2\mu''-4
r^2\mu'\nu'+r^2\mu'^2+4r\mu'-4r\nu'-4e^{\nu}+4)+\frac{2e^{-\nu}\nu'
}{r^3}(2r^2\mu''
\\\nonumber
&-&4r^2\mu'\nu'+r^2\mu'^2+4r\mu'-4r\nu'-4e^{\nu}+4)-\frac{e^{-\nu}
\nu''}{2 r^2}(2r^2\mu''-4r^2\mu'\nu'
\\\nonumber
&+&r^2\mu'^2+4r\mu'-4r\nu'-4 e^{\nu}+4)-\frac{e^{-\nu}\nu'}{r^2}(2
r^2\mu'''-4r^2\mu''\nu'+8r\mu''
\\\nonumber
&-&4r^2\mu'\nu''-8r\mu'\nu'+2r\mu'^2+4
\mu'+2r^2\mu'\mu''-4r\nu''-4e^{\nu}\nu'-4\nu')
\\\nonumber
&-&\frac{2e^{-\nu}}{r^3}
(2r^2\mu'''-4r^2\mu''\nu'+8r\mu''-4r^2\mu'\nu''-8r\mu'\nu'+2r
\mu'^2+4\mu'
\\\nonumber
&+&2r^2\mu'\mu''-4r\nu''-4e^{\nu}\nu'-4\nu')
+\frac{e^{-\nu}}{2r^2}(2r^2\mu^{(4)}-4r^2\mu'''\nu'+12r \mu'''
\\\nonumber
&-&8r^2
\mu''\nu''-16r\mu''\nu'+2r^2\mu''^2+12\mu''-4r^2\nu'''\mu'-16r\mu'
\nu''-8\mu'\nu'
\\\nonumber
&+&2\mu'^2+2r^2\mu'''\mu'+8r\mu'\mu''-4r\nu'''-4
e^{\nu}\nu''-8\nu''-4e^{\nu}\nu'^2),
\\\nonumber
G'&=&\frac{2e^{-2\nu}}{r^{3}}(2(e^{\nu}-1)\mu'^{2}+(6-e^{\nu})
r\mu'\nu'^{2} +2(e^{\nu}-1)(2\mu''-\mu''')+r\mu'\nu''
\\\nonumber
&\times&(e^{\nu}-3)-2(e^{\nu}-1)\mu''
+\nu'\{2(3-e^{\nu})\mu'+(3e^{\nu}-7)r\mu''+(e^{\nu}-2)r\mu'^{2}\}),
\\\nonumber
G''&=&\frac{2e^{-2\nu}}{r^{4}}\bigg[\mu'^{2}{6-6e^{\nu}
+(e^{\nu}-2)r^{2}\nu''}+(e^{\nu}-12)r^{2}\mu'\nu'^{3}-2\big\{\mu''
\big\{6(e^{\nu}-1)
\\\nonumber
&-&(2e^{\nu}-5)r^{2}\nu''+(e^{\nu}-1)r^{2}\mu''^{2}+(e^{\nu}-1)r(2\mu'''-4\mu''')
\big\}\big\}+\nu'\big\{\mu'(6
\\\nonumber
&\times&(e^{\nu}-3)+4(e^{\nu}-2)r^{2}\mu''-3(e^{\nu}-6)r^{2}\nu'')
-4(e^{\nu}-2)r\mu'^{2}+r(\mu'''r
\\\nonumber
&\times&(5e^{\nu}-11)-4(3e^{\nu}-7)\mu'')\big\}-4r\nu'^{2}(e^{\nu}-5)r\mu''-(e^{\nu}-6)\mu'+r\mu'^{2}
\\\nonumber
&\times&(e^{\nu}-4)+r\mu'\big\{8\mu''(e^{\nu}-1)-4(e^{\nu}-3)\nu''+r\big((e^{\nu}-3)\nu'''
\\\nonumber
&-&2\mu'''(e^{\nu}-1) \big)\big\}\bigg].
\end{eqnarray}

To investigate the existence of viable cosmic structures, it is
necessary to apply some specific constraints on matter named as
energy conditions. These conditions consist of a set of inequalities
that impose limitations on the stress-energy tensor which governs
the behavior of matter and energy in the presence of gravity. There
are several energy conditions, each of which places different
constraints on the stress-energy tensor as
\begin{itemize}
\item Null energy constraint
\begin{eqnarray}\nonumber
0\leq P_{r}+\varrho, \quad 0\leq P_{\bot}+\varrho.
\end{eqnarray}
\item Dominant energy constraint
\begin{eqnarray}\nonumber
0\leq \varrho-P_{r}, \quad 0\leq \varrho-P_{\bot}.
\end{eqnarray}
\item Weak energy constraint
\begin{eqnarray}\nonumber
0\leq P_{r}+\varrho,\quad 0\leq P_{\bot}+\varrho, \quad 0\leq
\varrho.
\end{eqnarray}
\item Strong energy constraint
\begin{eqnarray}\nonumber
0\leq P_{r}+\varrho, \quad 0\leq P_{\bot}+\varrho, \quad 0\leq
P_{r}+2P_{\bot}+\varrho.
\end{eqnarray}
\end{itemize}
These energy bounds are significant in determining the presence of
cosmic structures. Moreover, these conditions considerably impact
the existence of traversable $\mathcal{WH}$ geometry and other
hypothetical objects in spacetime. The viable $\mathcal{WH}$
structure must violate these conditions.

We consider $q(r)=\chi r^{3}$ \cite{35}, where $\chi$ is an
arbitrary constant. We choose $\chi=2$ for our convenience in all
the graphs. The graphical behavior of energy bounds is analyzed in
Figure \textbf{2}. The plots in the upper panel indicate that the
behavior of $\varrho+P_{r}$ is positive but negative behavior of
$\varrho+P_{\bot}$ implies that the null energy condition is
violated. The middle part shows that the dominant energy constraint
is violated due to the negative behavior of $\varrho-P_{r}$. The
behavior of $\varrho$ and $\varrho+P_{r}+2P_{\bot}$ is also
negative, which violates the weak and strong energy conditions,
respectively. These graphs manifest that fluid parameters violate
the energy conditions, especially the violation of the null energy
condition and negative behavior of energy density provides viable
traversable WH structure in the $f(R,G)$ gravity model.
\begin{figure}
\epsfig{file=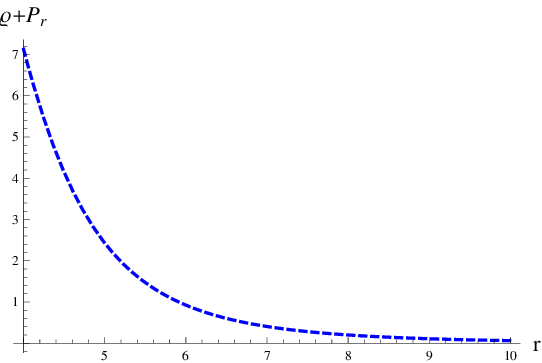,width=.5\linewidth}
\epsfig{file=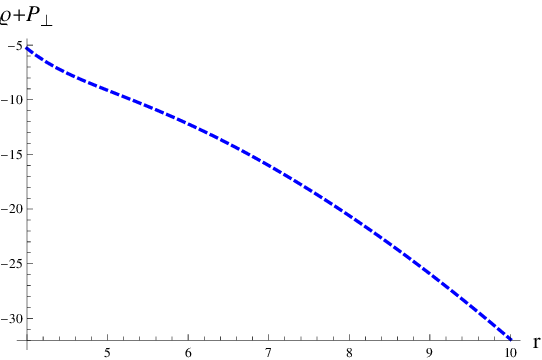,width=.5\linewidth}
\epsfig{file=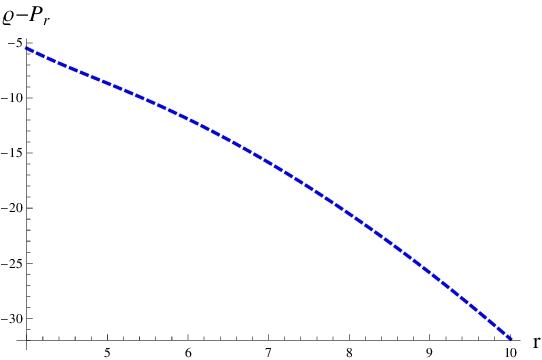,width=.5\linewidth}
\epsfig{file=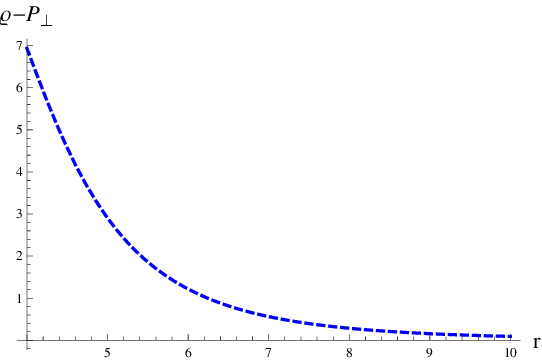,width=.5\linewidth}
\epsfig{file=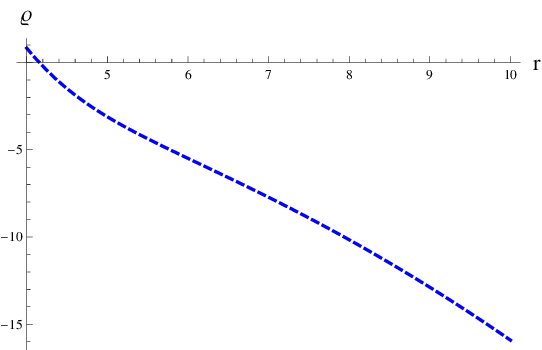,width=.5\linewidth}
\epsfig{file=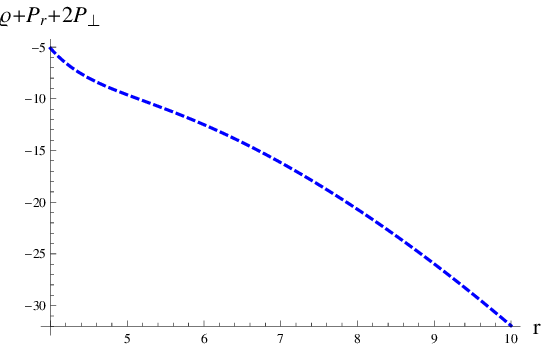,width=.5\linewidth}\caption{Plots of energy
conditions corresponding to radial coordinate for $\lambda=0.0005$,
$\alpha=20$, $\beta=80$ and $n=2$.}
\end{figure}

\section{Stable Wormhole Solutions}

Stability analysis investigates the constraints under which cosmic
structures remain stable against various oscillation modes. The
stability of a $\mathcal{WH}$ is determined by its geometry and the
properties of matter that creates and maintains it. Different
approaches examine $\mathcal{WH}$ stability. One approach involves
considering the sufficient exotic matter for generating the
necessary gravitational forces to keep the $\mathcal{WH}$ open.
Another approach involves considering the causality condition and
the Herrera cracking method. However, stability analysis of a
$\mathcal{WH}$ remains a complex and ongoing area of research. It
involves a comprehensive study of the behavior of matter around and
within the $\mathcal{WH}$. Understanding the concept of a stable
$\mathcal{WH}$ deepens our comprehension of the fundamental nature
of spacetime and opens up new possibilities for technological
advancements and novel forms of space travel. In the following, we
focus on analyzing the stability of viable and traversable
$\mathcal{WH}$ solutions by examining the \emph{causality condition}
and \emph{Herrera cracking}.

\subsection{Causality Condition}

The stability analysis of a $\mathcal{WH}$ can be examined through
the causality condition, which ensures that no signals or
information can travel faster than the speed of light. This
condition states that the time-like interval between any two events
in spacetime must always be greater than or equal to zero. In other
words, no signal can travel faster than the speed of light and
causality is preserved. In the case of a $\mathcal{WH}$, we need to
check whether the $\mathcal{WH}$ allows for the existence of closed
time-like curves. If a $\mathcal{WH}$ allows for the existence of
closed time-like curves, then causality would be violated and the
$\mathcal{WH}$ would be unstable. Therefore, we need to rule out the
possibility of such curves to ensure the stability of a
$\mathcal{WH}$. This can be done by analyzing the geometry of the
$\mathcal{WH}$ and ensuring that the time-like interval between any
two events is always greater than or equal to zero. The causality
condition states that the speed of sound components defined as
\begin{eqnarray}\nonumber
\nu_{sr}^{2}=\frac{dP_{r}}{d\varrho}, \quad \nu_{st}^{2}
=\frac{dP_{\bot}}{d\varrho},
\end{eqnarray}
must be confined in the interval of [0,1] for stable structures
\cite{35a}. The expressions of $\nu_{sr}^{2}$ and $\nu_{st}^{2}$ are
given in Appendix \textbf{A}. Figure \textbf{3} shows that static
spherically symmetric solutions are in the stable state for specific
values of model parameters as they fulfill the necessary constraint
for the $f(R,G)$ gravity model. Thus, this modified theory gives
stable traversable $\mathcal{WH}$ solutions.
\begin{figure}
\epsfig{file=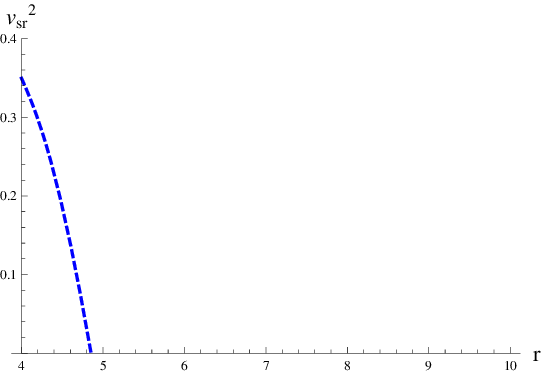,width=.5\linewidth}
\epsfig{file=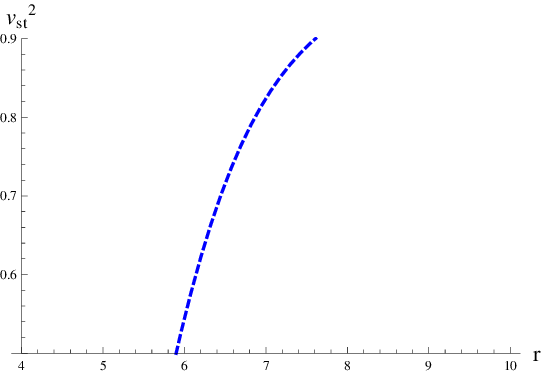,width=.5\linewidth} \caption{Behavior of
$\nu_{sr}^{2}$ and $\nu_{st}^{2}$ corresponding to radial coordinate
for for $\lambda=0.0005$, $\alpha=20$, $\beta=80$ and $n=2$.}
\end{figure}

\subsection{Herrera Cracking Technique}

The Herrera cracking technique is a mathematical tool to study the
stability of solutions. The stability of a $\mathcal{WH}$ solution
is essential because it determines whether the $\mathcal{WH}$ can
exist for a long time without collapsing or becoming unstable. The
stability of a $\mathcal{WH}$ solution can be determined by
analyzing the behavior of the cracking condition $(0\leq\mid
\nu_{sr}^{2}-\nu_{st}^{2}\mid\leq1)$ \cite{35b}. If the cracking
condition is violated, then the $\mathcal{WH}$ is unstable and will
collapse. On the other hand, if the cracking condition is satisfied,
then the $\mathcal{WH}$ is stable and can exist for a long time.
Hence, the Herrera cracking technique is a powerful mathematical
tool that analyze the stability of $\mathcal{WH}$ solutions. The
graphical behavior of $|\nu_{sr}^{2}-\nu_{st}^{2}|$ is given in
Figure \textbf{4} which demonstrates that physically stable static
spherically symmetric $\mathcal{WH}$ solutions exist corresponding
to particular values of model parameters.
\begin{figure}\center
\epsfig{file=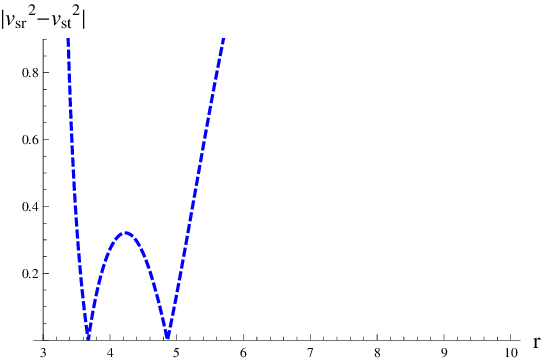,width=.5\linewidth} \caption{Behavior of
$|\nu_{sr}^{2}-\nu_{st}^{2}|$ corresponding to radial coordinate for
$\lambda=0.0005$, $\alpha=20$, $\beta=80$ and $n=2$.}
\end{figure}

\section{Final Remarks}

Various methods have been used in literature to obtain viable
$\mathcal{WH}$ structures. One of them is to formulate a shape
function through different methods and the other is to examine the
behavior of energy constraints by considering different
$\mathcal{WSF}$s. In the present article, we have studied the viable
$\mathcal{WH}$ geometry through embedding class-I in $f(R,G)$
gravity. We have considered a specific model of this modified theory
to find the exact solutions of static spherical spacetime. We have
examined the viability of traversable $\mathcal{WH}$ geometry
through the energy conditions. The obtained results are summarized
as follows

The newly developed shape function through the Karmarkar condition
satisfies all the necessary conditions which ensure the presence of
physically viable $\mathcal{WH}$ geometry (Figure \textbf{1}). We
have shown that the fluid parameters violate the energy conditions
especially the violation of the null energy condition for specific
values of $\lambda$, $\alpha$ $\beta$ and $n$ gives the existence of
exotic matter at the $\mathcal{WH}$ throat (Figure \textbf{2}).
Thus, we have obtained the viable traversable $\mathcal{WH}$
geometry in the framework of $f(R,G)$ theory. The stability limits,
i.e., $\nu_{sr}^{2}$ and $\nu_{st}^{2}\in[0,1]$(causality condition)
and $0<|\nu_{sr}^{2}- \nu_{st}^{2}|<1$ (Herrera cracking) are
satisfied which ensures the existence of physically viable and
stable $\mathcal{WH}$ geometry.

Fayyaz and Shamir \cite{36} used the Karmarkar condition to examine
physically viable traversable $\mathcal{WH}$ geometry in the
existence of exotic matter in $\mathcal{GR}$. For this purpose, they
developed a shape with the help of the embedding class-I technique
and found that the developed shape function satisfied the required
conditions. They obtained viable $\mathcal{WH}$ structures but did
not check their stable state. Later, the same authors \cite{37}
extended their work in $f(R)$ theory and obtained viable
$\mathcal{WH}$ solutions with negligible exotic matter. They did not
checked the stability analysis in this modified framework. Recently,
Sharif and Fatima \cite{33} generalized this work for $f(R,T)$
theory and obtained viable as well as stable $\mathcal{WH}$
solutions for minimum radius. Furthermore, the viable traversable
$\mathcal{WH}$ solutions were found in the context of $f(R,G)$
gravity due to the violation of energy conditions, indicating the
presence of exotic matter at $\mathcal{WH}$ throat. This implies
that the physically viable traversable $\mathcal{WH}$ solutions via
the Karmarkar condition exist in $f(R,G)$ theory.

Traversable $\mathcal{WH}$s are a fascinating concept in theoretical
physics and their study in the context of $f(R,G)$ theory has both
practical applications and theoretical implications. The presence of
traversable $\mathcal{WH}$ raises the possibility of time travel
because they can allow for closed timelike curves. In $f(R,G)$
gravitational framework, the existence of traversable
$\mathcal{WH}$s challenge our understanding of spacetime geometry.
It suggests that the modification of gravity from the standard
Einstein-Hilbert action produces exotic structures like wormholes,
opening new avenues for exploring the fundamental laws of physics.
The study of traversable $\mathcal{WH}$s reveals insights into the
effects of modified gravity on the universe's large-scale structure.
This helps refine our cosmic evolution models and the universe's
fate. Viable traversable $\mathcal{WH}$s in $f(R,G)$ theory leads to
a deeper understanding of the energy conditions that govern
spacetime and their compatibility with exotic matter, challenging
the classical energy conditions. Thus, the practical applications of
traversable $\mathcal{WH}$s in $f(R,G)$ theory revolutionize space
travel and cosmology, while their theoretical implications reshape
our understanding of fundamental physics, including gravity,
spacetime and the nature of the universe.

In this paper, we have investigated the viability and stability of
traversable $\mathcal{WH}$ solutions in the framework of $f(R,G)$
theory. The primary focus of our study is to explore the theoretical
aspects of the $\mathcal{WH}$ solutions and their behavior in the
presence of charge and modified terms. Traversable $\mathcal{WH}$s
have fascinated theoretical physics as they connect distant regions
of spacetime. Identifying viable and stable traversable
$\mathcal{WH}$ solutions in $f(R,G)$ theory is not merely a
theoretical exercise. It has profound implications for our
understanding of the universe's fundamental laws. Firstly, our
results contribute to the growing evidence suggesting that
traversable $\mathcal{WH}$s can exist in modified gravity theories
such as $f(R,G)$ theory. This is significant because it challenges
the conventional belief that such structures can only be
theoretically viable in the $\mathcal{GR}$ framework. By
demonstrating their existence in this modified framework, we have
opened new possibilities for studying and exploring these exotic
structures. This implies they exist in broader theoretical
frameworks than previously thought. Secondly, the stability of these
$\mathcal{WH}$ solutions is crucial for their practical utility. Our
analysis shows that these solutions satisfy causality conditions and
do not exhibit instability through Herrera cracking, indicating that
they could serve as stable conduits for travel between distant
regions of the universe. Hence, our research advances the
understanding of traversable $\mathcal{WH}$s in a modified framework
and provides a theoretical foundation for further exploration.

\vspace{0.25cm}

\section*{Appendix A}
\renewcommand{\theequation}{A\arabic{equation}}
\setcounter{equation}{0} \vspace{0.25cm}

The value of sound speed components against radial and tangential
components is given by
\begin{eqnarray}\nonumber
v_{sr}^{2}&=&\bigg[-\frac{e^{-2\nu}}{r^3}\bigg\{e^{2\nu}r^2(R+\lambda
R^{2}+\alpha G^{n}+\beta G\ln(G))-e^{2\nu}G r^2(\alpha n
G^{n-1}+\beta
\\\nonumber
&+&\beta\ln(G))+2\lambda R'e^{\nu}r(\mu'r+4)-e^{\nu}(1+2\lambda
R)(e^{\nu}R r^2-2\mu'r+2e^{\nu}-2)
\\\nonumber
&-&4\mu'G'(e^{\nu}-3)(\alpha(n^{2}-n) G^{n-2}+\beta G^{-1})
+\frac{e^{2\nu}q^2}{4\pi
r^2}\bigg\}-\frac{\nu'e^{-2\nu}}{r^2}\bigg\{e^{2\nu}r^2(R
\\\nonumber
&+&\lambda R^{2}+\alpha G^{n}+\beta G\ln(G))-e^{2\nu}(\alpha n
G^{n-1}+\beta+\beta\ln(G))Gr^2+2\lambda R'e^{\nu}r
\\\nonumber
&\times&(\mu'r+4)-e^{\nu}(1+2\lambda
R)(e^{\nu}Rr^2-2\mu'r+2e^{\nu}-2)-4\mu'(e^{\nu}-3)(\beta G^{-1}
\\\nonumber
&+&\alpha(n^{2}-n)G^{n-2})G'+\frac{e^{2 \nu}q^2}{4\pi
r^2}\bigg\}+\frac{e^{-2\nu}}{2r^2}\bigg\{2\nu'(R+\lambda
R^{2}+\alpha G^{n}+\beta G\ln(G))
\\\nonumber
&\times&e^{2\nu}r^2-2e^{2\nu}(\alpha nG^{n-1}+\beta+\beta\ln(G))G
\nu'r^2+2\lambda R'e^{\nu}\mu'\nu'r^2-e^{2\nu}Gr^2G'
\\\nonumber
&\times&(\alpha (n^{2}-n) G^{n-2}+\beta
G^{-1})+e^{2\nu}r^2\big\{R'+\beta G'\ln(G)+\beta G'+\alpha
nG^{n-1}G'
\\\nonumber
&+&2\lambda RR'\big\}+2e^{2\nu}r(R+\lambda R^{2}+\alpha G^{n}+\beta
G\ln(G))-2e^{2\nu} -e^{2\nu}(\alpha nG^{n-1}+\beta
\\\nonumber
&+&\beta\ln(G))G'r^2+2\lambda R'e^{\nu}\mu''r^2+2\lambda
R''e^{\nu}\mu'r^2(\alpha nG^{n-1}+\beta+\beta\ln(G))Gr
\\\nonumber
&+&4\lambda R'e^{\nu}r(\mu'+2\nu')+8\lambda
e^{\nu}(rR''+R')-(2\lambda R'e^{\nu}+e^{\nu}\nu'(1+2\lambda R))
\\\nonumber
&\times&(e^{\nu}Rr^2-2\mu'r+2e^{\nu}-2)-4G'(e^{\nu}\mu'\nu'+e^{\nu}\mu''-3\mu'')
(\alpha (n^{2}-n) G^{n-2}
\\\nonumber
&+&\beta G^{-1})-(1+2\lambda R)(e^{\nu}R\nu'
r^2+e^{\nu}R'r^2+2e^{\nu}Rr-2\mu''r-2\mu'+2e^{\nu}\nu')
\\\nonumber
&\times&e^{\nu}-4\mu'\mathcal{A}(e^{\nu}-3)+\frac{e^{2\nu}q^2
\nu'}{2\pi r^2}+\frac{e^{2\nu}qq'}{2\pi r^2}-\frac{e^{2\nu}q^2}{2\pi
r^3}\bigg\}\bigg]\bigg[-\frac{e^{-2 \nu}}{r^3}\bigg\{-e^{2\nu}r^2
\\\nonumber
&\times&(R+\lambda R^{2}+\alpha nG^{n-1}+\beta
G\ln(G))+e^{2\nu}(\alpha n G^{n-1}+\beta+\beta\ln(G))Gr^2+2
\\\nonumber
&\times&(R'\nu'-2R'')\lambda e^{\nu}r^2-8\lambda
R'e^{\nu}r+e^{\nu}(1+2\lambda R)(e^{\nu}Rr^2+2\nu'r+2e^{\nu}-2)
\\\nonumber
&-&(4e^{\nu}-12)(\alpha (n^{2}-n) G^{n-2}+\beta
G^{-1})\nu'G'+8\mathcal{A}(e^{\nu}-1)-\frac{e^{2\nu}q^2}{4\pi
r^2}\bigg\}-\frac{1}{r^2}
\\\nonumber
&\times&e^{-2 \nu}\nu'\bigg[-e^{2\nu}r^2(R+\beta G\ln(G)+\lambda
R^{2}+\alpha G^{n})+2\lambda e^{\nu}r^2(R'\nu'-2R'')
\\\nonumber
&-&8\lambda R'e^{\nu}r+e^{2\nu}(\alpha n
G^{n-1}+\beta+\beta\ln(G))Gr^2+e^{\nu}(1+2\lambda
R)(e^{\nu}Rr^2+2\nu'r
\\\nonumber
&+&2e^{\nu}-2)-4(e^{\nu}-3)\nu'G'(\alpha(n^{2}-n) G^{n-2}+\beta
G^{-1})+8\mathcal{A}(e^{\nu}-1)-\frac{e^{2\nu}q^2}{4\pi r^2}\bigg]
\\\nonumber
&+&\frac{e^{-2\nu}}{2 r^2} \bigg[2\lambda
R'e^{\nu}\nu'^2r^2-2e^{2\nu}\nu'r^2(R+\lambda R^{2}+\alpha
G^{n}+\beta G\ln(G))+2e^{2\nu}(\beta\ln(G)
\\\nonumber
&+&\alpha n G^{n-1}+\beta)G\nu'r^2+e^{2\nu}Gr^2G'(\beta
G^{-1}+\alpha(n^{2}-n)G^{n-2})-e^{2\nu}r^2(2\lambda RR'
\\\nonumber
&+&R'+\alpha n G^{n-1}G'+\beta G'\ln(G)+\beta G')+e^{2\nu}(\alpha
nG^{n-1}+\beta+\beta\ln(G))G'r^2
\\\nonumber
&-&2\lambda r^2e^{\nu}(R''\nu'-R'\nu''+2R''')-2e^{2\nu}r(R+\lambda
R^{2}+\alpha G^{n}+\beta G\ln(G))+2e^{2\nu}
\\\nonumber
&\times&(\alpha nG^{n-1}+\beta+\beta\ln(G))Gr-4\lambda
e^{\nu}(R'\nu'r+4 R''r+2R')+2\lambda R'e^{\nu}(e^{\nu}Rr^2
\\\nonumber
&+&2\nu'r+2e^{\nu}-2)+e^{\nu}\nu'(1+2\lambda R)(e^{\nu}Rr^2+2
\nu'r+2e^{\nu}-2)-4(e^{\nu}\nu'^2+e^{\nu}\nu''
\\\nonumber
&-&3\nu'')(\beta G^{-1}+\alpha (n^{2}-n) G^{n-2})G'
+e^{\nu}(1+2\lambda R)(e^{\nu}R\nu' r^2+e^{\nu}R'r^2+2e^{\nu}Rr
\\\nonumber
&+&2\nu''r+2e^{\nu}\nu'+2\nu')+4\nu'\mathcal{A}(e^{\nu}+3)+\mathcal{A}'(8
e^{\nu}-1)+\frac{e^{2\nu}q^2}{2\pi r^3}-\frac{e^{2\nu}q^2\nu'}{2\pi
r^2}
\\\nonumber
&-&\frac{e^{2\nu}q q'}{2\pi r^2}\bigg]\bigg]^{-1},
\\\nonumber
v_{st}^{2}&=&\bigg[-\frac{e^{-2\nu}}{4r^2}\bigg\{e^{\nu}r(1+2\lambda
R)\mu'^2+2e^{2\nu}r(R+\lambda R^{2}+\alpha G^{n}+\beta
G\ln(G))-2e^{2\nu}r
\\\nonumber
&\times&(\alpha
nG^{n-1}+\beta+\beta\ln(G))G-2e^{2\nu}(rR-\mu')(1+2\lambda
R)+8\lambda R'e^{\nu}(1+2r\mu')
\\\nonumber
&-&e^{\nu}(2\nu'+r\mu'\nu'+2r\mu'')(1+2\lambda R)-4\lambda
R'e^{\nu}r\nu'+4G'(\mu'^2-3\mu'\nu'+2\mu'')
\\\nonumber
&\times& (\alpha (n^{2}-n) G^{n-2}+\beta G^{-1})+8\lambda
R''e^{\nu}r+8\mu'\mathcal{A}-\frac{e^{2\nu}q^2}{2\pi
r^3}\bigg\}-\frac{1}{2 r}\bigg[e^{-2\nu}\nu'
\\\nonumber
&\times&\bigg\{e^{\nu}(r\mu'^2-2e^{\nu}rR+2\mu'-2
\nu'-r\mu'\nu'+2r\mu'')(1+2\lambda R)+2 (R+\lambda R^{2}
\\\nonumber
&+&\alpha G^{n}+\beta G\ln(G))e^{2\nu}r-2e^{2\nu}r(\alpha
nG^{n-1}+\beta+\beta\ln(G))G+4\lambda(2+r\mu'
\\\nonumber
&-&r\nu')R'e^{\nu}+4G'(\mu'^2-3\mu'\nu'+2\mu'')(\alpha (n^{2}-n)
G^{n-2}+\beta G^{-1})+8\lambda R''e^{\nu}r
\\\nonumber
&+&8\mu'\mathcal{A}-\frac{e^{2\nu}q^2}{2\pi
r^3}\bigg\}\bigg]+\frac{e^{-2\nu}}{4 r}\bigg\{e^{\nu}(1+2\lambda
R)\big\{\mu'^2-2\nu'^2-r\mu'\nu'^2 -2\nu''+\mu'\nu'
\\\nonumber
&+&r\mu'^2\nu'-r\mu'\nu''+2r\mu'\mu''+r\nu'\mu''+2r
\mu'''-4e^{\nu}rR\nu'-2e^{\nu}R-2e^{\nu}rR'
\\\nonumber
&+&4\mu''\big\}+R'\lambda e^{\nu}\big\{-4r\nu'^2+8r\mu''-4
e^{\nu}rR+8\mu'-4r\nu''+2r\mu'\nu'+2r\mu'^2\big\}
\\\nonumber
&+&\lambda
e^{\nu}(4r\mu'+4r\nu'+16)R''-(2e^{2\nu}G-4e^{2\nu}rG\nu'-2e^{2\nu}rG')(\alpha
nG^{n-1}+\beta
\\\nonumber
&+&\beta\ln(G))+(2e^{2\nu}+4e^{2\nu}r\nu')(R+\lambda R^{2}+\alpha
nG^{n-1}+\beta G\ln(G))+2e^{2\nu}r(R'
\\\nonumber
&+&2\lambda RR'+\alpha n G^{n-1}G'+\beta G'\ln(G)+\beta G')+2G'
(4\mu'\mu''-e^{2\nu}rG-6\nu'\mu''
\\\nonumber
&-&6\mu'\nu''+4\mu''')(\alpha(n^{2}-n) G^{n-2}+\beta
G^{-1})+4\mathcal{A}(\mu'^2-3\mu'\nu'+4\mu'')+8\mu'\mathcal{A}'
\\\nonumber
&+&8\lambda R'''e^{\nu}r-\frac{e^{2\nu}\nu'q^2}{\pi
r^3}+\frac{3e^{2\nu}q^2}{2\pi r^4} -\frac{e^{2\nu}q'q}{\pi
r^3}\bigg\}\bigg]\bigg[-\frac{e^{-2\nu}}{r^3}\bigg\{-e^{2 \nu}r^2
(R+\lambda R^{2}
\\\nonumber
&+&\alpha G^{n}+\beta G\ln(G))+e^{2\nu}(\alpha n
G^{n-1}+\beta+\beta\ln(G))Gr^2+2\lambda R're^{\nu}(\nu'r-4)
\\\nonumber
&-&4\lambda R''e^{\nu}r^2+e^{\nu}(1+2\lambda R)(e^{\nu}
Rr^2+2\nu'r+2e^{\nu}-2)+4G'(3\nu'-e^{\nu}\nu')
\\\nonumber
&\times&(\alpha (n^{2}-n) G^{n-2}+\beta
G^{-1})+8\mathcal{A}(e^{\nu}-1)-\frac{e^{2\nu} q^2}{4\pi
r^2}\bigg\}-\frac{e^{-2\nu}\nu'}{r^2}\bigg\{-e^{2 \nu}r^2
\\\nonumber
&\times&(R+\lambda R^{2}+\alpha G^{n}+\beta G\ln(G))
+e^{2\nu}(\alpha nG^{n-1}+\beta+\beta\ln(G))Gr^2
\\\nonumber
&+& 2\lambda R'e^{\nu}r(\nu'r-4)-4\lambda R''e^{\nu}r^2+
e^{\nu}(1+2\lambda R)(e^{\nu} Rr^2+2\nu'r+2e^{\nu}-2)
\\\nonumber
&+&4G'\nu'(3-e^{\nu})(\alpha (n^{2}-n) G^{n-2}+\beta G^{-1})
+8\mathcal{A}(e^{\nu}-1)-\frac{e^{2\nu} q^2}{4\pi
r^2}\bigg\}+\frac{e^{-2\nu}}{2r^2}
\\\nonumber
&\times&\bigg\{2\lambda
R'e^{\nu}(\nu'^2r^2+\nu''r^2-4+e^{\nu}Rr^2+2\nu'r+2e^{\nu}-2-2\nu'r)
-16\lambda R''e^{\nu}r
\\\nonumber
&-&2e^{2\nu}r(\nu'r+1)(R+\lambda R^{2}+\alpha G^{n}+\beta
G\ln(G))+e^{\nu}r^2(2e^{\nu}G\nu'+G')(\beta
\\\nonumber
&+&\alpha nG^{n-1}+\beta\ln(G))+e^{2\nu}Gr^2G'(\alpha (n^{2}-n)
G^{n-2}+\beta G^{-1})-e^{2\nu}r^2(R'
\\\nonumber
&+&2\lambda RR'+\alpha n G^{n-1}G'+\beta G'\ln(G)+\beta G')-2\lambda
R''e^{\nu}\nu'r^2-4\lambda R'''e^{\nu}r^2
\\\nonumber
&+&2e^{2\nu}(\alpha nG^{n-1}+\beta+\beta\ln(G)) Gr
+e^{\nu}\nu'(1+2\lambda R)(e^{\nu}Rr^2+2\nu'r+2e^{\nu}
\\\nonumber
&-&2)-4e^{\nu}G'(\nu'^2+\nu''+3e^{-\nu}\nu'')(\alpha (n^{2}-n)
G^{n-2}+\beta G^{-1})+e^{\nu}(1+2\lambda R)
\\\nonumber
&\times&(e^{\nu}R\nu'r^2+e^{\nu}R'r^2+2
e^{\nu}Rr+2\nu''r+2e^{\nu}\nu'+2\nu')
+4\mathcal{A}(e^{\nu}\nu'+3\nu')
\\\nonumber
&+&8\mathcal{A}'(e^{\nu}-1)+\frac{e^{2\nu}q^2}{2\pi
r^3}-\frac{e^{2\nu}q^2\nu'}{2\pi r^2}-\frac{e^{2\nu}qq'}{2\pi
r^2}\big\}\bigg]^{-1},
\end{eqnarray}
where
\begin{eqnarray}\nonumber
\mathcal{A}&=&\alpha(n-1)nG^{n-2}G''+\beta G''G^{-1}+\alpha(n-2)
(n-1)nG^{n-3}G'^2-\beta G'^2G^{-2},
\\\nonumber
\mathcal{A}'&=&\alpha(n-1)nG^{n-2}G'''+\beta
G'''G^{-1}+\alpha(n-3)(n-2)(n-1)nG^{n-4}G'^3
\\\nonumber
&+&2\beta G'^3G^{-3}+3\alpha(n-2)(n-1)nG^{n-3}G'G''-3\beta G'
G''G^{-2}.
\end{eqnarray}
\textbf{Data Availability Statement:} No new data was created or
analysed in this study.

\end{document}